\documentclass[letterpaper]{article} 
\usepackage{aaai2026}  
\usepackage{times}  
\usepackage{helvet}  
\usepackage{courier}  
\usepackage[hyphens]{url}  
\usepackage{graphicx} 
\usepackage{natbib}  
\usepackage{caption} 
\usepackage{algorithm}
\usepackage{algorithmic}

\usepackage{xcolor}
\usepackage{amssymb}
\usepackage{booktabs}
\usepackage{multirow}
\usepackage{makecell}
\usepackage{pifont}
\usepackage{amsmath}
\usepackage{amsthm}
\usepackage{subcaption}
\usepackage{bm}
\usepackage{microtype}

\usepackage[utf8]{inputenc} 
\usepackage{amsfonts}       
\usepackage{nicefrac}       
\usepackage{colortbl}

\usepackage{xr}

\usepackage{newfloat}
\usepackage{listings}
\DeclareCaptionStyle{ruled}{labelfont=normalfont,labelsep=colon,strut=off} 
\floatstyle{ruled}
\newfloat{listing}{tb}{lst}{}
\floatname{listing}{Listing}
\title{MLLM-based Discovery of Intrinsic Coordinates and Governing Equations from High-Dimensional Data}
\author{
    Ruikun Li,~
    Yan Lu\thanks{Corresponding author (luyan@pjlab.org.cn).},~
    Shixiang Tang,~
    Biqing Qi,~
    Wanli Ouyang
}
\affiliations{
    \textsuperscript{\rm 1} Shanghai Artificial Intelligence Laboratory
}

\nocopyright

\begin{document}

\maketitle

\begin{abstract}
Discovering governing equations from scientific data is crucial for understanding the evolution of systems, and is typically framed as a search problem within a candidate equation space. However, the high-dimensional nature of dynamical systems leads to an exponentially expanding equation space, making the search process extremely challenging. The visual perception and pre-trained scientific knowledge of multimodal large language models (MLLM) hold promise for providing effective navigation in high-dimensional equation spaces.
In this paper, we propose a zero-shot method based on MLLM for automatically discovering physical coordinates and governing equations from high-dimensional data. Specifically, we design a series of enhanced visual prompts for MLLM to enhance its spatial perception. In addition, MLLM’s domain knowledge is employed to navigate the search process within the equation space.
Quantitative and qualitative evaluations on two representative types of systems demonstrate that the proposed method effectively discovers the physical coordinates and equations from both simulated and real experimental data, with long-term extrapolation accuracy improved by approximately 26.96\% compared to the baseline.
\end{abstract}

\section{Introduction}

Nonlinear dynamics are prevalent across numerous fields in science and engineering, such as astrophysics, fluid mechanics, and materials science~\cite{tenachi2023deep, wang2019symbolic, li2025predicting, li2025predicting3}. 
Identifying the symbolic equations governing the dynamics from data is crucial for understanding and predicting such systems, a task known as symbolic regression~\cite{langley1981data, schmidt2009distilling,cornelio2023combining}. 
Traditional symbolic regression approaches typically assume that the physical coordinates are of low dimensionality and focus on developing more efficient and accurate search algorithms~\cite{biggio2021neural, shojaee2023transformer, la2021contemporary}. 
However, real-world scientific scenarios are often hindered by the high-dimensional nature of observational data, making it difficult to navigate the exponentially expanding symbolic space, even though a significant portion of the high-dimensional variables may be redundant~\cite{chen2022automated, pope2021intrinsic}. 
Reasoning physical knowledge from high-dimensional observational data to facilitate predictive understanding~\cite{liu2025physgen, duan2022survey} poses a novel and interdisciplinary challenge.

From Kepler’s discovery of planetary motion laws to the most advanced symbolic regression techniques, the discovery of equations has historically relied on scientists manually processing observational data to obtain compact physical quantities in suitable coordinate systems~\cite{chen2022automated, luan2022distilling}. 
However, automatically discovering both physical coordinates and the governing equations on them from high-dimensional data is a challenging problem.
With advancements in deep learning, especially visual models~\cite{krizhevsky2012imagenet, he2016deep, dosovitskiy2020image}, discovering physical laws from high-dimensional data, like videos, has become more feasible.
Unsupervised motion video prediction with neural networks integrated with parameterized physical modules (e.g., neural ODE~\cite{chen2018neural}) allows for implicit learning of dynamics from high-dimensional observations~\cite{guen2020disentangling, wu2023disentangling, higgins2021symetric,hofherr2023neural, li2025predicting2, bounou2021online}.
Another approach is to embed physical engines as inductive biases into the model pipeline, compelling the neural network to infer physical quantities that satisfy the governing physical laws from video streams~\cite{kadambi2023incorporating, yang2022learning, fotiadis2023disentangled, kandukuri2022physical, jaques2019physics}.
However, these methods separate the discovery of coordinates and equations, presuming known dynamics equations or physics engines before identifying physical quantities.
Recent efforts have succeeded in jointly discovering physical coordinates and equations~\cite{champion2019data, luan2022distilling, zhang2024vision}, yet they still rely on prior knowledge of equation forms and state dimensions, limiting their generalization to unfamiliar systems.

High-dimensional data challenges traditional optimization methods like Bayesian optimization due to the lack of well-defined quantitative evaluations for intrinsic coordinates and governing equations.
Large language models (LLMs) excel in problem-solving by leveraging pretrained knowledge and enhancing prompts for scientific discovery across various domains.~\cite{boiko2023autonomous, m2024augmenting, kang2024chatmof}.
Multimodal large language models (MLLMs) with visual encoders achieve spatial understanding akin to human visual intelligence.~\cite{wu2025dettoolchain}.
Recent studies~\cite{liu2025physgen} show MLLMs can deduce physical properties and motion patterns from images without prior knowledge, revealing their potential in high-dimensional data.
Moreover, most symbolic regression methods, based on genetic programming, search equation spaces through mutation and recombination. By leveraging the pretraining knowledge of LLM to guide this search, equation discovery can be more efficient and avoid suboptimal solutions~\cite{shojaee2024llm, ma2024llm}. 
Thus, we can anticipate that a well-designed reasoning framework and visual prompts could harness MLLMs for visual perception and scientific discovery, enabling automatic identification of joint physical coordinates and governing equations.

In this work, we propose a Video Equation Reasoning framework (VER) based on multimodal large language models to automatically discover physical coordinates and their corresponding governing equations from high-dimensional data. VER first introduces a set of enhanced visual prompts designed for MLLMs to uncover the intrinsic physical coordinate space of observed data. In the equation reasoning module, we employ a hypothesis-assessment-iteration reasoning chain to refine existing symbolic regression methods for equation discovery. By integrating coordinate discovery and equation reasoning, VER is the first to explore the potential of MLLMs in equation discovery from high-dimensional data. Quantitative and qualitative evaluations on two types of representative spatiotemporal dynamical systems demonstrate that the proposed method accurately discovers the physical coordinates and governing equations from both simulated and experimental data. The extrapolation accuracy of the discovered equations improves by approximately 26.96\% on average compared to the baseline.

Our contributions can be summarized as follows:
\begin{itemize}
    \item We design a series of visual locating tools and feedback prompts for MLLMs to discover low-dimensional physical coordinates in complex spatiotemporal dynamics.
    \item We propose a hypothesis generation and optimization evaluation strategy that leverages MLLMs’ pretrained scientific knowledge, significantly enhancing the performance of existing methods in equation reasoning.
    \item Extensive experiments on representative dynamical systems demonstrates that VER surpasses existing methods by a significant margin and is capable of extracting physical insights from high-dimensional observations of real-world systems.
\end{itemize}
\section{Problem Fomulation}

In the task of inferring governing equations for high-dimensional data, the goal is to find a compact and accurate symbolic expression. Here, we consider a class of video-like systems governed by low-dimensional coordinates, that is, systems whose dynamics evolve on a low-dimensional manifold~\cite{champion2019data}. Given the dataset $\mathcal{D}=\{x_i\}^n_{i=1}$, where $x_i \in \mathbb{R}^D$ is the $i$-th observed state (i.e., a frame of video), we aim to uncover the underlying mathematical relationships such that $\frac{dx}{dt}=f(x)$. This fundamentally relies on an effective coordinate system $\mathcal{Z}$, in which the dynamics have a simple low-dimensional representation $\frac{dz}{dt}=\hat{f}(z)$. Therefore, this type of problem consists of two coupled subproblems:
\begin{itemize}
    \item \textbf{Discovering physical coordinates} is to learning a pair of bidirectional transformations $\phi : \mathbb{R}^D \rightarrow \mathbb{R}^d$ and $\psi : \mathbb{R}^d \rightarrow \mathbb{R}^D$, which map the equation discovery in high-dimensional space $\frac{dx}{dt}=f(x)$ to the intrinsic coordinate system $\frac{dz}{dt}=g(\phi(x))=g(z)$.
    \item \textbf{Reasoning governing equations} means inferring a closed-form expression for the dynamics $g$ governing the system based on its evolution trajectory in the low-dimensional coordinate system, also known as symbolic regression.
\end{itemize}
The discovered physical coordinates directly influence the governing equation, such as the number of variables and the simplicity of the equation. A good equation should not only accurately fit the observed data points but also exhibit strong generalization capability on unseen data.

\section{Methodology}

\begin{figure*}[!ht]
    \centering
    \includegraphics[width=0.98\linewidth]{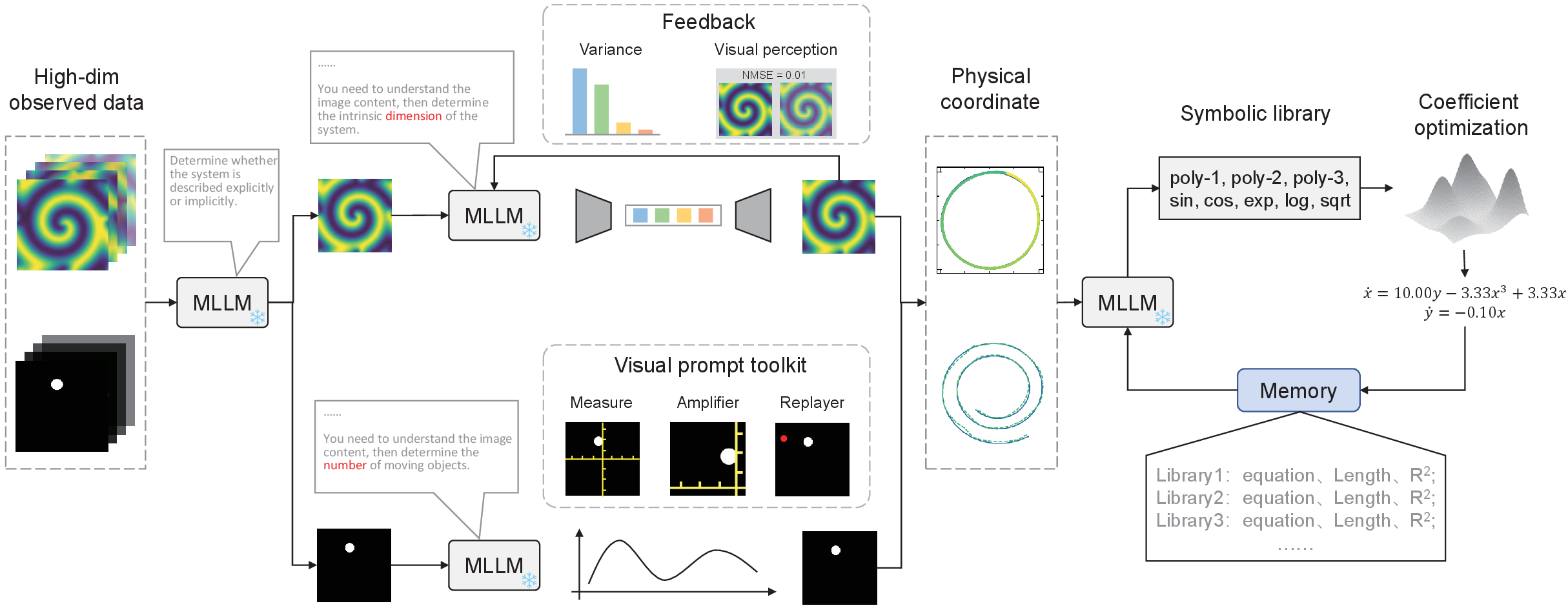}
    \vspace{-0.1cm}
    \caption{Overall pipeline of proposed video equation reasoning framework.}
    \label{fig:framework}
\end{figure*}

While state-of-the-art MLLMs possess powerful visual understanding and reasoning capabilities, our approach aims to incrementally unlock their ability to infer dynamical equations from high-dimensional observational data (video modality). We propose a comprehensive reasoning framework, VER, which includes the entire Chain-of-Thought (CoT) to discover physical coordinates and reason governing equations from visual inputs, as shown in Figure~\ref{fig:framework}.

\subsection{Discovering physical coordinates}

We consider two types of high-dimensional systems. The first type is the pixel coordinate system, where physical laws are presented in the form of Cartesian coordinates of rigidly moving objects \cite{luan2022distilling, zhang2024vision, liu2025physgen}. MLLM needs to identify the intrinsic physical coordinate system by locating the trajectory of the target object's motion through a sequence of coordinates. The second type is the more complex latent coordinate system, whose spatiotemporal evolution is governed by latent low-dimensional variables, presenting periodic or quasiperiodic oscillatory patterns \cite{champion2019data, rudy2017data}. We sample a small number of frames from the data and ask MLLM to determine their type in order to initiate different locating tools.

\subsubsection{Detection for Pixel Coordinate}

Let $\mathbb{L}$ and $\mathbb{I}$ denote a set of language characters (prompts) and a frame of continuous video, respectively. Given a set of queries $q=\{x_l,x_i || x_l \in \mathbb{L}, x_i \in \mathbb{I}\}$ corresponding to the motion video, we aim to obtain the corresponding physical coordinate sequence $Z=\{z\}^n_{i=1}$ with the help of a frozen MLLM, working as $\phi$. The coordinate sequence is retrieved through a string extractor $e$, which fetches substrings within specific delimiters from the historical outputs. MLLM serially traverses each query and maintains a history, including locating outputs from previous messages. Considering that visual prompting is more effective than language prompting for locating tasks, we design a locating toolkit, $\mathrm{LocateTool}$, to be applied to each query of MLLM, thereby enhancing MLLM's spatial understanding through augmented visual prompts. We design three types of locating tools as enhanced visual prompts, which serve to make judgments before and after MLLM's locating.

\textbf{Spatial Measurement} overlays a coordinate system with linear graduations and gridlines on the original image, providing a clearer reference for target localization. The graduations simplify the localization task, allowing MLLM to read the coordinates of the target rather than making predictions. The gridlines help MLLM further align the target with the coordinate graduations.

\textbf{Regional Amplifier} first calls MLLM to detect the quadrant where the target is located, then crops the area. By zooming into the local region of the target's quadrant, MLLM observes the spatial relationship between the target and the coordinate system at close range. To avoid misidentification, MLLM is additionally called to confirm whether the cropped area contains the target object as a secondary verification.

\textbf{Plot Replayer} displays MLLM's initial locating results as markers (e.g., red dots) on the original frame and calls MLLM to compare the visual discrepancy between the target object and the detection location for error correction. As a feedback iteration step in the reasoning chain, the replayer is used after the initial locating of each frame. However, for the sake of simplicity in the formula, we have incorporated it into the $\mathrm{LocateTool}$ functionality in Algorithm 1 in appendix.

After serially detecting the state of each frame, we introduce a filter tool to smooth the physical coordinate trajectory in order to enhance the continuity of motion between frames. Specifically, we designed a feedback correction module, $\mathrm{Filter}$, based on the Savitzky-Golay filter~\cite{savitzky1964smoothing}. We plot the recognized trajectory as a visual prompt and ask MLLM to determine the window length $h$ and polynomial order $p$ based on the evolution trend and periodicity of the trajectory. Each filtering result is visualized alongside the originally recognized trajectory as the visual prompt for the next iteration, until MLLM determines that the smoothing has preserved the evolution trend while maintaining inter-frame coordinate continuity. The pseudocode for serially extracting the physical coordinate sequence from the high-dimensional motion video is shown in Algorithm 1 in appendix.

\subsubsection{Detection for Latent Coordinate}

The task of discovering physical coordinates in the latent coordinate system is no longer visual target localization, but rather the encoding of intrinsic variables that capture complex spatiotemporal evolution patterns~\cite{chen2022automated}. Here, intrinsic variables refer to the minimal set of variables that describe all the information of a dynamic system~\cite{li2023learning}. Although prior work has developed manifold learning methods such as Koopman operator theory and delayed embedding~\cite{lusch2018deep, wu2024predicting}, these models heavily rely on prior assumptions about the system's intrinsic dimensions (i.e., the number of intrinsic variables).

In this work, we use an autoencoder as an auxiliary tool to design a feedback-iteration reasoning chain for MLLM to automatically determine the intrinsic dimensions of unknown systems. Specifically, in each iteration of MLLM, we record a pair of encoder-decoder $\phi_d$ and $\psi_d$ with a bottleneck dimension $d$ and perform self-supervised reconstruction on each frame of the observed evolution trajectory $x_i \in \mathbb{R}^D$, obtaining the intrinsic variable sequence $Z=\{z\}^n_{i=1}$. We maintain an experience buffer to store the reconstruction error corresponding to the bottleneck dimension $d$ and the reconstruction samples as the text and visual prompts for the next iteration. Furthermore, we compute the variance explained ratio $\{\frac{\lambda_i}{\sum^d_{j=1}\lambda_j}\}^d_{i=1}$ of the $d$-dimensional intrinsic variables, where $\lambda_i$ is the eigenvalue of the $i$-th principal component, to enhance the prompt. The differences in the magnitude of variance contribution rates help MLLM determine whether redundant dimensions exist and decide the bottleneck dimension for the next iteration. Finally, the most suitable bottleneck dimension $d$, as determined by MLLM, along with the corresponding encoder $\phi_d$ and decoder $\psi_d$, are input into the governing equation reasoning module.

\subsection{Reasoning governing equations}

We adopt the sparse identification method for nonlinear dynamics (SINDy) \cite{brunton2016discovering} as the starting point for our approach. Inspired by the fact that latent governing equations typically contain only a finite number of terms, SINDy identifies the equations by optimizing the coefficients of predefined symbolic terms (e.g. $sin$, $exp$, and $x^2$). Specifically, SINDy defines a symbolic library $\Theta(z)=[\theta_1(z),\theta_2(z),...,\theta_k(z)] \in \mathbb{R}^{d \times k}$, where $\theta_i$ represents a candidate symbolic term. The symbolic library is linearly combined to form a deterministic system $\frac{dz}{dt}=\Theta(z)\Xi$, where $\Xi=[\xi_1,\xi_2,...,\xi_k] \in \mathbb{R}^{k \times 1}$ is the coefficient to be optimized for each symbolic term. Given the coordinate sequence $Z=\{z_i\}^n_{i=1}$ and its derivative sequence $\dot{Z}=\{\dot{z_i}\}^n_{i=1}$, SINDy minimizes the loss function 
\begin{equation}\label{loss:sindy}
    L_{SINDy} = \frac{1}{n} \sum^n_{i=1}||\dot{z_i} - \Theta(z_i)\Xi|| + \eta |\Xi|
\end{equation}
to ensure both fitting accuracy and sparsity.

\subsubsection{Hypothesis generation}

Previous SINDy-based works have succeeded \cite{chen2021physics, gao2022autonomous}, but rely on a high-quality predefined symbolic library, often needing prior system knowledge to manually filter out unnecessary terms. 
To address systems without prior knowledge, we design a hypothesis generation module to mine the physical knowledge pre-learned by MLLM and automatically select potential symbolic terms. 
The hypothesis generation step utilizes the frozen MLLM to propose diverse and promising candidate symbolic terms. 
At each iteration $t$, MLLM uses replay samples from the experience pool (Sec.~\ref{sec:experience}) as text and visual prompts, suggesting the candidate symbolic library $\Theta_t$.

\subsubsection{Hypothesis Assessment}

After generating the candidate symbolic library hypothesis $\Theta_t$, we utilize the detected physical coordinate data for optimization and scoring. This process involves the stochastic gradient optimization of the coefficient matrix $\Xi_t$ and evaluating its fit. Specifically, for the pixel coordinate system, we train using the loss function from Equation~\ref{loss:sindy}. For the latent coordinate system, we jointly fine-tune the encoder $\phi_d$, decoder $\psi_d$, and $\Xi_t$ to ensure that the intrinsic physical coordinates align with the physical equation. The loss function 
\begin{equation}
\begin{split}
    L_{AE-SINDy} ={}& \frac{1}{n} \sum^n_{i=1} ||x_i-\psi_d(x_i)||^2_2 \\
    & + ||\nabla_{x_i}(z_i)\dot{x}_i-\Theta_t(z_i)\Xi_t||^2_2 \\
    & + \eta |\Xi|
\end{split}
\end{equation}
coordinates the overall performance of self-supervised reconstruction and equation discovery.

We employ the $R^2$ metric as a quantitative measure of how well the equation fits the data. Additionally, we evaluate the number $length_t$ of symbolic terms in the equation to quantify its simplicity. In symbolic regression tasks, the goal of maintaining equation simplicity is to avoid overfitting to the limited observational data. At this stage, the candidate symbolic library hypothesis $\Theta_t$ is assessed based on both the $R^2$ metric and the equation length to determine its $fitness_t$.
Thus, the equation discovery process is divided into two steps: (i) constructing the symbolic library, and (ii) optimizing based on stochastic gradient using PyTorch~\cite{kingma2014adam}. 
This process aligns with the philosophy of human scientific discovery~\cite{ma2024llm, guods2024DS, huang2024mlagentbench}, where the MLLM makes decisions on the discrete components of the SINDy symbolic library based on pretrained knowledge and prior experiences in prompts, akin to the empirical hypotheses made by human scientists during scientific discovery.

\subsubsection{Experience management}\label{sec:experience}

To improve the search efficiency in symbolic space and avoid local optima, we employ an experience management mechanism to enhance prompts during hypothesis generation. 
For each iteration, we construct a sample tuple $(\Theta_t,\Xi_t,fitness_t)$ and store it in the experience pool, thus maintaining a diverse and high-quality candidate symbolic library. 
In the next hypothesis generation, we collect the most recent $m$ historical experiences as an enhanced prompt for MLLM to generate a more promising symbolic library, where $m$ represents the receptive field.
MLLM is encouraged to compare equations derived from different symbolic libraries to remove redundant symbolic terms that lead to overly complex and overfitting, while exploring other promising new symbolic terms.

We implement an early stopping mechanism to handle convergence scenarios before reaching the maximum iteration limit. Once the symbolic library suggested by MLLM reaches a threshold of repetition with samples in the experience pool, the equation inference step will be terminated. At this point, MLLM will be asked to select the best governing equation from the experience pool, using fitting accuracy and simplicity as the criteria.
\section{Experiments}

We validate VER's advantages over baselines using two types of representative systems. For each system type, we collect simulated and real videos and use VER to extract physical insights. Additionally, we conduct robustness and ablation experiments to further discuss the benefits of each MLLM component of VER compared to existing methods.

\subsection{Datasets}

We evaluate the proposed method in the dynamic systems of existing studies, including the pixel coordinate system formed by the motion of interested objects according to physical laws and the latent coordinate system described by nonlinear partial differential equations.

\begin{figure*}[!t]
    \centering
    \includegraphics[width=\linewidth]{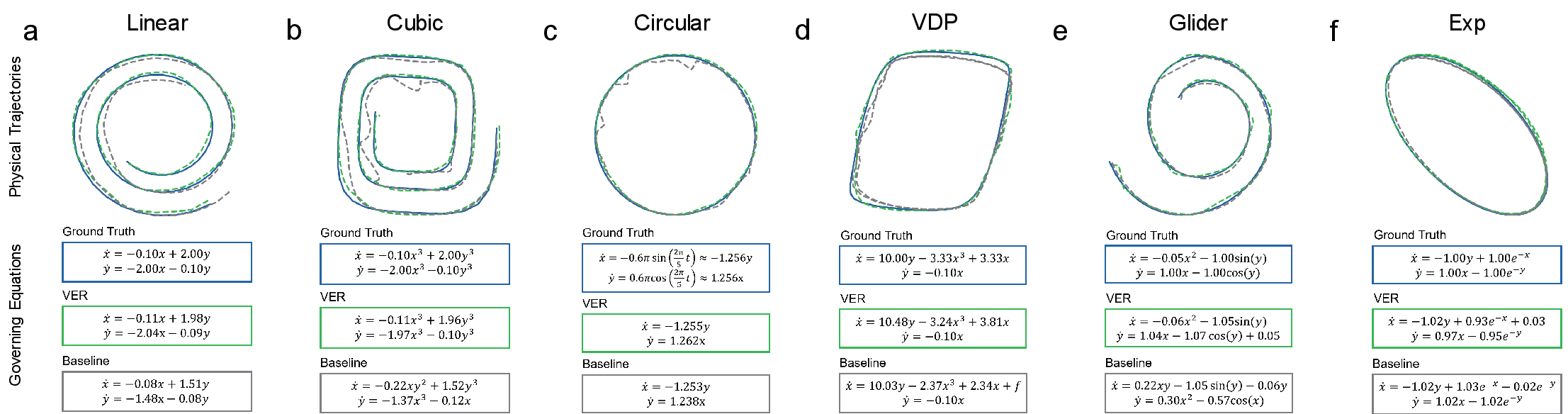}
    \caption{Reasoning results of pixel coordinate systems: blue line is ground truth; green and gray dashed lines are trajectories inferred by VER and the optimal baseline.}
    \label{fig:experiment1}
\end{figure*}

\textbf{Pixel coordinate system.} We validate the effectiveness of the proposed method on the pixel coordinate system using several dynamical systems shown in Figure~\ref{fig:experiment1}, including: \textit{Linear}, \textit{Cubic}, \textit{Circular}, \textit{Van Der Pol (VDP)}, \textit{Glider}, and \textit{Exp} equations. 
These equations include various characteristics such as linear terms, nonlinear terms, and significant differences in time scales~\cite{guckenheimer1980dynamics}, which together form an evaluation benchmark across different levels of difficulty.
The video data is generated by simulating physical trajectories, with details provided in the Appendix~\ref{app:data}.

\textbf{Latent coordinate system.} We select three representative reaction-diffusion equations and a real video as the latent coordinate system:
\begin{itemize}
    \item Lambda–Omega (LO) equation~\cite{champion2019data} generates spiral wave structures featuring two oscillating spatial modes.
    \item Brusselator (Bruss) equation~\cite{lopez2022gd} models the concentration dynamics in chemical reactions, where the system's trajectory eventually reaches a limit cycle attractor.
    \item Shallow-Water (Water) equation~\cite{takamoto2022pdebench}, derived from the compressible Navier-Stokes equations,describes fluid dynamics in shallow water scenarios, commonly applied in large-scale geophysical flows and tsunami simulations.
    \item Real video~\cite{Schmitt2024} visualizes the formation of Kármán vortex streets as water flows past a cylindrical obstacle with a Reynolds number of 171, using dye to highlight the vortex pattern.
\end{itemize}

We construct the dataset by solving the differential equations numerically. The details of the data generation are provided in Appendix~\ref{app:data}.

\subsection{Experimental Setup}

\textbf{Models.}
We utilize the GPT-4o~\cite{islam2024gpt} in our experiments to explore the performance of our proposed VER. We validate the improvements of VER over the basic GPT-4o in the ablation study.
The penalty coefficient $\eta$ in Equation~\ref{loss:sindy} is set to $0.01$ and we explore the performance of hyperparameter search in the appendix~\ref{app:adaptive_eta}.

\textbf{Baselines.}
Related works~\cite{luan2022distilling, udrescu2021symbolic} on discovering coordinates from pixel coordinate videos typically employ a combination of visual localization models and symbolic regression models. We adopt a recent work~\cite{luan2022distilling}, Video-SINDy, as the baseline method for the pixel coordinate system. In addition, we also replace the coordinate encoder in Video-SINDy with SAM~\cite{kirillov2023segment} as a zero-shot baseline, SAM-SINDy.
For the latent coordinate system, the most relevant work~\cite{champion2019data} constructs an end-to-end model combining an autoencoder and SINDy (AE-SINDy) to interpret low-dimensional patterns in high-dimensional data. Since it requires prior knowledge of the low-dimensional space dimensions, we ensure that its dimensions are consistent with those discovered by VER.

\subsection{System with Pixel Coordinate}

Figure~\ref{fig:experiment1} shows VER's reasoning results in the pixel coordinate system, including governing equations and inferred trajectories compared to the optimal baseline. VER uses only raw video input, without needing motion type or equation terms. VER's trajectories closely align with ground truth over time, demonstrating MLLM's spatial understanding through locating prompts, as supported by \cite{wu2025dettoolchain}. The inferred ODE terms perfectly match the ground truth, indicating VER's ability to uncover intrinsic dynamics in pixel space.

We assess if VER and baseline methods' discovered equations include all correct terms, count incorrect terms, and evaluate fitting accuracy ($R^2$). We numerically integrate discovered equations to predict trajectories, comparing them to real ones in Table~\ref{tab:pixel_quantative}. VER consistently outperforms baselines, especially in long-term predictions, where inaccuracies are magnified. This highlights VER's superior efficiency and accuracy over traditional visual symbolic regression methods. Additionally, we collect physical motion videos from public sources to confirm VER's reliability on real systems (Appendix~\ref{app:real_motion}).

\begin{table}[!ht]
\renewcommand{\arraystretch}{1.}
\centering
\setlength{\tabcolsep}{1pt}
\resizebox{\linewidth}{!}{%
\begin{tabular}{ccccccc}
\toprule
 & \multirow{2}{*}{Method} & \multicolumn{3}{c}{Equation} & \multicolumn{2}{c}{Prediction}  \\
 & & \makecell{Terms \\ Found} & \makecell{False \\ Positives} & $R^2$ & $R^2@100$ & $R^2@1000$  \\
\midrule
 \multirow{3}{*}{\rotatebox{90}{Linear}} & Video-SINDy & Yes & $0.91\pm0.34$ & $0.95\pm0.02$ & $0.90\pm0.03$ & $0.87\pm0.06$  \\
 & SAM-SINDy & Yes & $0.45\pm0.13$ & $0.90\pm0.03$ & $0.87\pm0.03$ & $0.35\pm0.13$  \\
 & VER & Yes & $\mathbf{0.38\pm0.22}$ & $\mathbf{0.96\pm0.03}$ & $\mathbf{0.97\pm0.02}$ & $\mathbf{0.96\pm0.02}$  \\
\midrule
 \multirow{3}{*}{\rotatebox{90}{Cubic}} & Video-SINDy & No & $3.52\pm1.12$ & $0.82\pm0.07$ & $0.68\pm0.18$ & $0.26\pm0.19$  \\
 & SAM-SINDy & No & $2.17\pm1.20$ & $0.76\pm0.09$ & $0.70\pm0.12$ & $0.22\pm0.18$  \\
 & VER & Yes & $\mathbf{1.41\pm0.73}$ & $\mathbf{0.95\pm0.03}$ & $\mathbf{0.80\pm0.16}$ & $\mathbf{0.62\pm0.28}$  \\
\midrule
 \multirow{3}{*}{\rotatebox{90}{Circular}} & Video-SINDy & Yes & $0.31\pm0.10$ & $0.98\pm0.01$ & $0.98\pm0.03$ & $0.90\pm0.10$  \\
 & SAM-SINDy & Yes & $0.12\pm0.03$ & $0.92\pm0.01$ & $0.99\pm0.01$ & $0.93\pm0.04$  \\
 & VER & Yes & \textbf{0} & $\mathbf{0.99\pm0.01}$ & $\mathbf{1.00\pm0.00}$ & $\mathbf{0.97\pm0.05}$  \\
\midrule
 \multirow{3}{*}{\rotatebox{90}{VDP}} & Video-SINDy & Yes & $2.31\pm0.65$ & $0.89\pm0.08$ & $0.82\pm0.06$ & $0.49\pm0.13$  \\
 & SAM-SINDy & Yes & $2.45\pm0.59$ & $0.88\pm0.03$ & $0.92\pm0.04$ & $0.67\pm0.08$  \\
 & VER & Yes & $\mathbf{0.80\pm0.40}$ & $\mathbf{0.97\pm0.01}$ & $\mathbf{0.94\pm0.02}$ & $\mathbf{0.73\pm0.07}$  \\
\midrule
 \multirow{3}{*}{\rotatebox{90}{EXP}} & Video-SINDy & Yes & $2.76\pm0.88$ & $0.93\pm0.08$ & $0.93\pm0.02$ & $0.90\pm0.03$  \\
 & SAM-SINDy & Yes & $3.21\pm1.17$ & $0.95\pm0.02$ & $0.96\pm0.01$ & $0.95\pm0.02$  \\
 & VER & Yes & $\mathbf{1.34\pm0.29}$ & $\mathbf{0.99\pm0.01}$ & $\mathbf{0.99\pm0.01}$ & $\mathbf{0.98\pm0.02}$  \\
\midrule
 \multirow{3}{*}{\rotatebox{90}{Glider}} & Video-SINDy & No & $2.18\pm0.49$ & $0.94\pm0.02$ & $0.98\pm0.01$ & $0.91\pm0.03$  \\
 & SAM-SINDy & No & $2.23\pm0.56$ & $0.85\pm0.04$ & $0.99\pm0.04$ & $0.32\pm0.27$  \\
 & VER & Yes & $\mathbf{0.84\pm0.17}$ & $\mathbf{0.98\pm0.01}$ & $\mathbf{0.99\pm0.01}$ & $\mathbf{0.92\pm0.02}$  \\
\bottomrule
\end{tabular}%
}
\caption{Average performance of pixel coordinate systems over 10 runs with varying seeds. $R^2@n$ indicates $R^2$ for $n$ steps. Best results are in bold.}
\label{tab:pixel_quantative}
\end{table}

\begin{figure}[!ht]
  \centering
  \includegraphics[width=1\linewidth]{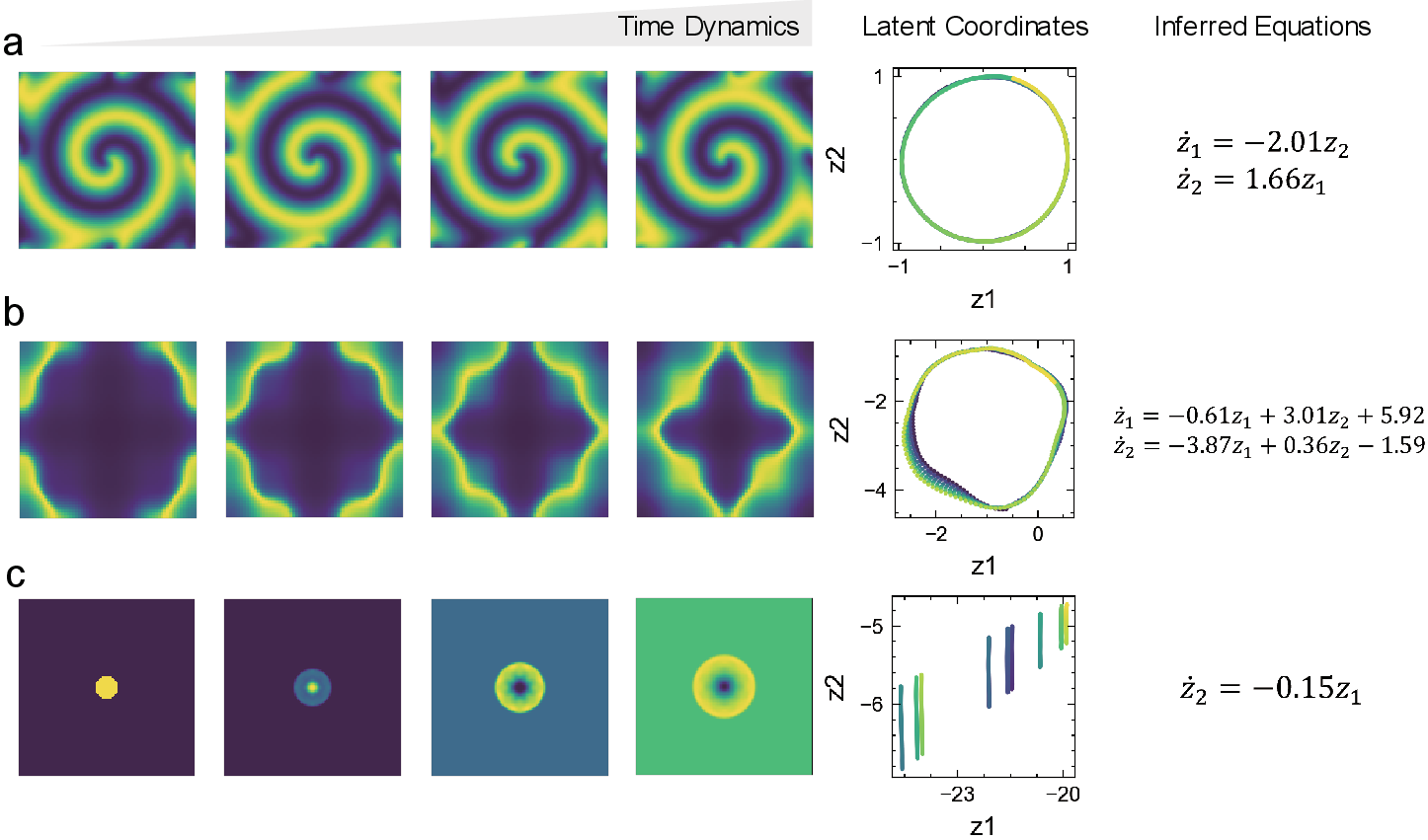}
  \caption{Reasoning results of the latent coordinate system (a) LO, (b) Bruss, and (c) Water. The trajectories show the low-dimensional oscillatory patterns found by VER.}
  \label{fig:experiment2}
\end{figure}

\begin{figure}[!ht]
  \centering
  \includegraphics[width=1\linewidth]{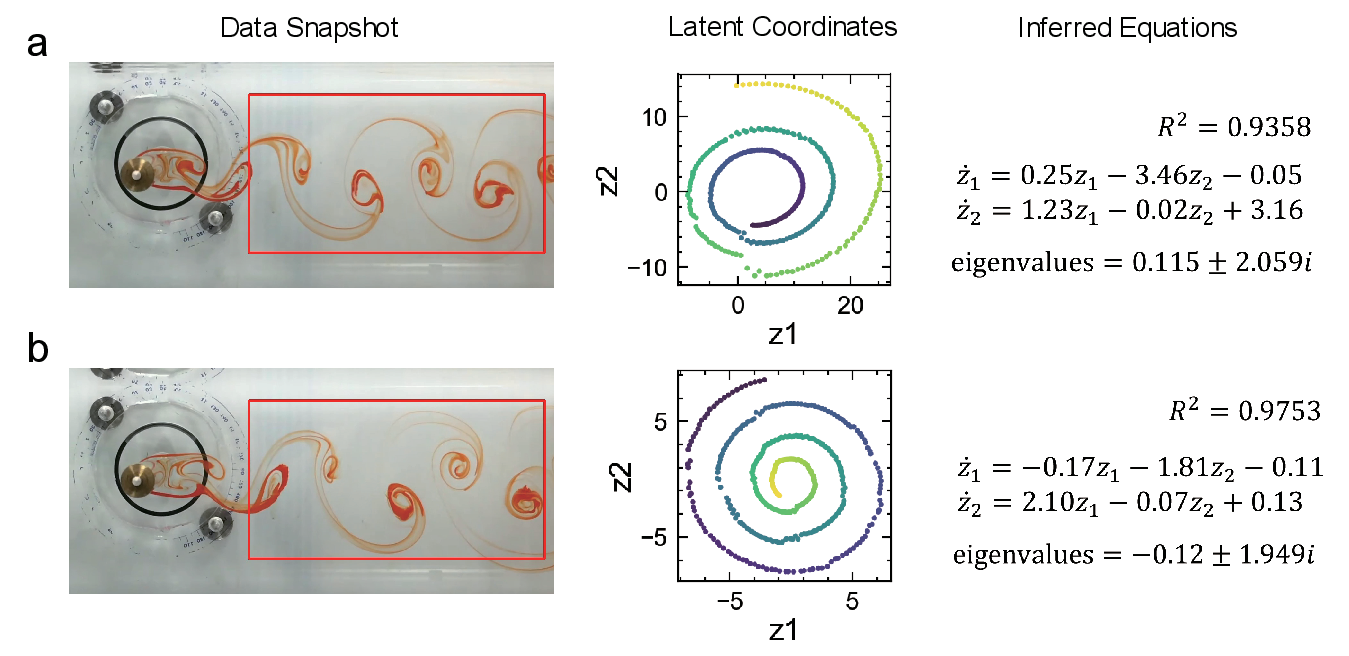}
  \caption{Kármán street video reasoning results: raw data yields vortex trajectories in (a) and (b). The red-box region's flow pattern is projected into 2D latent space.}
  \label{fig:experiment3}
\end{figure}

\subsection{System with Latent Coordinate}

The reasoning results of VER in the latent coordinate system are presented in Figure~\ref{fig:experiment2}, including the zero-shot inferred latent coordinate space and governing equations. VER takes observation sequences in the form of multi-channel image modalities as input and locates the intrinsic dimension of high-dimensional nonlinear partial differential equations. The results in Figure~\ref{fig:experiment2} show that, despite the high nonlinearity of the original equations, VER reveals the linear approximation of oscillatory patterns in the latent space spanned by the intrinsic dimensions, which is consistent with the conclusions of previous works \cite{champion2019data}.
Notably, in the case of the Water system, observed sequences evolved from different initial conditions are identified as trajectories with consistent dynamics in the latent space. While VER determines that two latent coordinates are required to capture the oscillatory patterns in the observation data, the results of the equation inference module indicate that the system's dynamics are actually governed by a one-dimensional linear ODE. The second dimension is only used to distinguish coordinate region differences caused by varying initial conditions.

To evaluate the physical insights extracted by VER from noisy real-world systems, we extract two datasets from recorded turbulence videos, each corresponding to a fixed region over two different time intervals, as illustrated in Figure~\ref{fig:experiment3}. Fluctuations in incoming flow velocity or direction, as well as surface roughness of the obstacle, lead to irregular vortex shedding, resulting in the gradual breakdown of the periodic structure of the vortex street. 
VER successfully identifies the linear dynamics equations of oscillatory modes underlying the high-dimensional turbulent flow, which manifest as a limit cycle with a changing radius. 
The eigenvalues of two linear equations are $0.115\pm2.059i$ and $-0.120\pm1.949i$. The positive (negative) real parts indicate that the system state diverges from (converges to) the equilibrium over time, consistent with the spiral process shown in Figure~\ref{fig:experiment2}. This indicates that in the first video, the vortex shedding around the cylinder exhibits an accelerating detachment trend. In the second video, the speed of vortex detachment gradually slows down. In addition, the similar imaginary parts indicate that the vortex shedding periods in two videos are close.

\subsection{Robustness}

We add Gaussian noise with varying strengths $\sigma$ (standard deviation, mean of $0$) to the observations, which biases derivative estimation and affects the accuracy and simplicity of the discovered equations.
Table~\ref{tab:noise} shows the performance of all methods on the latent coordinate system. VER exhibits significantly greater robustness to noise compared to the baseline.
This is because MLLM, guided by prompts, balances accuracy and equation simplicity during identification, avoiding coefficient estimation bias (overfitting) for irrelevant symbolic terms caused by noise disturbances.

\subsection{Ablation Study}

\begin{figure}[!ht]
  \centering
  \includegraphics[width=1\linewidth]{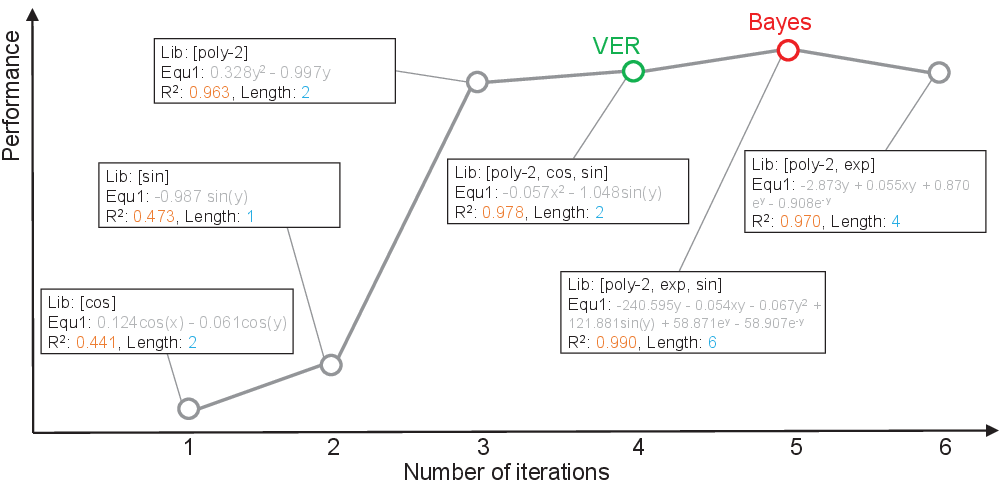}
  \caption{Comparison of VER and Bayesian optimization on \textit{Glider} system.}
  \label{fig:llm_vs_bayes}
\end{figure}

We conduct ablation studies to validate the contributions of VER's replaceable components to MLLM. 
As shown in Table~\ref{tab:ablation}, we evaluate several ablated versions on linear and cubic systems. 
Without our reasoning framework, the base MLLM (GPT-4o) fails to discover accurate equations from raw video, leading to large trajectory prediction errors. 
This confirms VER unlocks the MLLM's intrinsic reasoning for this task, rather than relying on potential data leakage. 
Ablating key components like the enhance visual prompts or the symbolic refiner (VER-SINDy) also lead to significant performance degradation or more redundant equation terms. 
Thus, VER's architecture is crucial for discovering symbolic equations from high-dimensional data. 
Further ablations on parameter selection (Appendix~\ref{app:bayes_ablation}) and symbolic libraries (Fig.~\ref{fig:llm_vs_bayes}) confirm the MLLM's superiority over traditional Bayesian methods. 
Finally, to test dependency on the backbone MLLM, we swap to Gemini-1.5-flash~\cite{team2024gemini}. 
Although this weaker model reduces performance, it still successfully identifies the correct equation terms.

\begin{table}[!ht]
\renewcommand{\arraystretch}{1.}
\centering
\setlength{\tabcolsep}{1pt}
\resizebox{\linewidth}{!}{%
\begin{tabular}{ccccccc}
\toprule
 & \multirow{2}{*}{Method} & \multicolumn{3}{c}{Equation} & \multicolumn{2}{c}{Prediction}  \\
 & & \makecell{Terms \\ Found} & \makecell{False \\ Positives} & $R^2$ & $R^2@100$ & $R^2@1000$  \\
\midrule
 \multirow{6}{*}{\rotatebox{90}{Linear}} & GPT-4o & Yes & $2.40\pm1.36$ & $0.25\pm0.23$ & $0.30\pm0.27$ & $0.07\pm0.38$  \\
 & w/o measure & No & $1.02\pm0.54$ & $0.82\pm0.06$ & $0.81\pm0.05$ & $0.70\pm0.09$ \\
 & w/o amplifier & Yes & $0.64\pm0.31$ & $0.91\pm0.03$ & $0.90\pm0.03$ & $0.88\pm0.04$ \\
 & w/o replayer & Yes & $0.51\pm0.18$ & $0.92\pm0.01$ & $0.93\pm0.02$ & $0.91\pm0.02$ \\
 & VER-SINDy & Yes & $1.25\pm0.65$ & $\mathbf{0.97\pm0.02}$ & $0.85\pm0.07$ & $0.83\pm0.07$  \\
 & VER(Gemini) & Yes & $0.58\pm0.23$ & $0.91\pm0.02$ & $0.92\pm0.03$ & $0.89\pm0.05$\\
 & VER & Yes & $\mathbf{0.38\pm0.22}$ & $0.96\pm0.03$ & $\mathbf{0.97\pm0.02}$ & $\mathbf{0.96\pm0.02}$  \\
\midrule
 \multirow{6}{*}{\rotatebox{90}{Cubic}} & GPT-4o & No & $4.40\pm1.39$ & $0.12\pm0.27$ & $0.11\pm0.25$ & $0.06\pm0.28$  \\
 & w/o measure & No & $2.49\pm0.82$ & $0.87\pm0.09$ & $0.74\pm0.15$ & $0.30\pm0.22$\\
 & w/o amplifier & No & $1.78\pm0.62$ & $0.91\pm0.05$ & $0.76\pm0.17$ & $0.49\pm0.26$\\
 & w/o replayer & Yes & $1.67\pm0.56$ & $0.93\pm0.04$ & $0.75\pm0.14$ & $0.56\pm0.23$\\
 & VER-SINDy & No & $2.85\pm0.93$ & $0.88\pm0.07$ & $0.71\pm0.17$ & $0.32\pm0.26$  \\
 & VER(Gemini) & Yes & $1.96\pm0.74$ & $0.90\pm0.07$ & $0.74\pm0.19$ & $0.41\pm0.25$\\
 & VER & Yes & $\mathbf{1.41\pm0.73}$ & $\mathbf{0.95\pm0.03}$ & $\mathbf{0.80\pm0.16}$ & $\mathbf{0.62\pm0.28}$  \\
\bottomrule
\end{tabular}%
}
\caption{Ablation studies on linear and cubic systems: 'w/o' means 'without'. VER-SINDy uses the standard SINDy algorithm instead of VER's inference module. VER(Gemini) uses Gemini-1.5-Flash as backbone MLLM.}
\label{tab:ablation}
\end{table}

\section{Related Work}

\subsection{Inferring physics from video}
Many studies focus on modeling video dynamics for predictive understanding, often using parameterized modules for implicit physics. Vincent et al.~\cite{guen2020disentangling} decoupled video content into PDE dynamics and unknowns, predicting dynamics via a parameterized PDE network. Wu et al.~\cite{wu2023disentangling} applied stochastic PDE dynamics for real-world randomness. Hofherr et al.~\cite{hofherr2023neural} and Bounou et al.~\cite{bounou2021online} used neural ODEs and Koopman operators as dynamic predictors for interpretable parameters. These methods limit interpretability by parameterizing dynamics. With known equations, Jaques et al.~\cite{jaques2019physics} used inverse graphics to capture physical quantities, while Kandukuri et al.~\cite{kandukuri2022physical} enforced rigid body mechanics via convolutional encoders to identify properties like mass and friction. Yang et al.~\cite{yang2022learning} and Fotiadis et al.~\cite{fotiadis2023disentangled} utilized encoder-decoder models for low-dimensional physical coordinates. Unlike our approach, these methods depend on prior knowledge or known equations, lacking the ability to infer physical coordinates and governing equations from unfamiliar videos.

\subsection{Multimodal large language models on vision tasks}
Recent research on multimodal large language models in vision tasks has demonstrated significant advancements. Wu et al.~\cite{wu2025dettoolchain} meticulously designed a reasoning chain to unlock the potential of MLLMs in object detection tasks. Liu et al.~\cite{liu2025physgen} further combined physics engines with MLLMs, enabling MLLMs to analyze motion targets and their physical properties from images to generate physics-grounded videos. Additionally, extensive research has focused on fine-tuning or reasoning chains to enhance MLLMs’ video understanding capabilities~\cite{zhou2024mlvu, li2024mvbench, zeng2024timesuite}. In contrast to these efforts, we pioneer the exploration of unlocking MLLMs’ ability to discover equations from videos through carefully designed tools and reasoning chains.

\subsection{Visual symbolic regression}
High-dimensional symbolic regression methods for videos is a novel scientific task that typically follows two paradigms: (1) object tracking and equation discovery, where trajectories are extracted and equations identified~\cite{luan2022distilling, zhang2024vision}; (2) encoder-decoder models for latent physical coordinates, mapping data into low-dimensional spaces for equation regression~\cite{udrescu2021symbolic, champion2019data}. Our approach uses MLLMs’ visual understanding and scientific knowledge for zero-shot discovery of both coordinates and equations, unifying these paradigms in one framework.
\section{Conclusion}

In this paper, we propose a video equation reasoning framework based on MLLMs to collaboratively discover physical coordinates and governing equations from high-dimensional observational data. We design an enhanced visual prompt toolkit to unify the discovery process of physical coordinates across two types of video-like systems. 
Extensive experiments demonstrate that the proposed method outperforms the baseline on two representative systems and provides a significant improvement to MLLMs. 
Additionally, the governing equations discovered by VER can quantitatively reveal the shedding period and evolving trends of vortices in real videos.
In summary, this work explores the potential of MLLM in discovering scientific knowledge from multimodal data.
\paragraph{Limitation} The performance of our framework is inherently linked to the quality of the upstream visual processing modules. Furthermore, its robustness under challenging visual conditions, such as severe occlusions or motion blur, warrants further exploration.
\bibliography{z_reference}
\clearpage
\clearpage
\appendix

\section*{Impact Statement}
This paper presents work whose goal is to advance the field of Machine Learning. There are many potential societal consequences of our work, none of which we feel must be specifically highlighted here.

\section{Data Generation} \label{app:data}

We synthesized simulated videos of pixel coordinate systems and latent coordinate systems using Python. First, the dynamical equations for each system are predefined, consistent with those in the main text. The evolution trajectories of each pixel coordinate system are simulated using the ODE solver function provided by Scipy. The 2D variables represent the x and y coordinates of moving objects in a 2D coordinate system. For the latent coordinate system described by partial differential equations, we adopt a discretized form of ordinary differential equations to numerically solve the system on a 2D spatial grid, generating data in a video-like modality. The size of each system is described in Table~\ref{tab:dataset}.

For the latent coordinate systems, we utilize partial differential equations (PDEs) to model complex dynamical behaviors. The following are the specific equations and parameter settings for each system:
\begin{itemize}
    \item Lambda–Omega (LO) equation
    
    This system generates spiral wave structures characterized by two oscillating spatial modes. The governing equations are
    \begin{equation}
    \begin{aligned}
        \frac{\partial u}{\partial t} &= (1-(u^2+v^2))u + \beta(u^2+v^2)v + d_1 \nabla^2 u, \\
        \frac{\partial v}{\partial t} &= -\beta(u^2+v^2)u + (1-(u^2+v^2))v + d_2 \nabla^2 v,
    \end{aligned}
    \end{equation}
    where the parameters are set to $d_1 = d_2 = 0.1$ and $\beta = 1$. These equations simulate the spatial dynamics observed in wave propagation phenomena.

    \item Brusselator (Bruss) equation

    The Brusselator models the reaction-diffusion process in chemical reactions, where the dynamic trajectory approaches a limit cycle attractor after initial transients. The equations are
    \begin{equation}
    \begin{aligned}
        \frac{\partial u}{\partial t} &= a - (1+b)u + vu^2 + d_1 \nabla^2 u, \\
        \frac{\partial v}{\partial t} &= bu - vu^2 + d_2 \nabla^2 v,
    \end{aligned}
    \end{equation}
    with diffusivities set as $d_1 = 1$, $d_2 = 0.1$ and reaction rates $a = 1$, $b = 3$. This system is well-suited for studying oscillatory chemical dynamics.
    
    \item Shallow-Water (Water) equation
    
    This system provides a framework for modeling free-surface flow problems such as tsunamis or flooding events. The equations are
    \begin{equation}
    \begin{aligned}
        \partial_t h + \nabla \cdot h \mathbf{u} &= 0, \\
        \partial_t h \mathbf{u} + \nabla \left( \mathbf{u}^2 h + \frac{1}{2} g_r h^2 \right) &= -g_r h \nabla b.
    \end{aligned}
    \end{equation}
    We use open-source simulation data from the related work~\cite{Schmitt2024}.
    
\end{itemize}

\section{Additional Results} \label{app:add_results}

\subsection{Real-world motion video} \label{app:real_motion}

\begin{figure*}[!ht]
    \centering
    \includegraphics[width=1.0\linewidth]{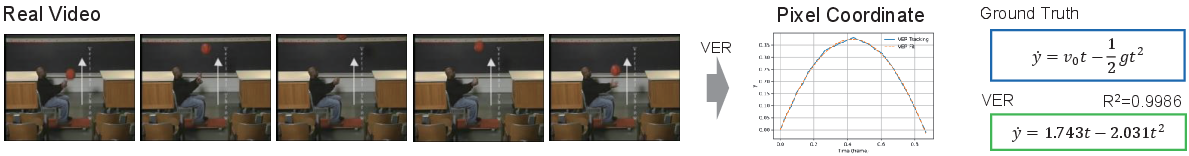}
    \vspace{-0.75cm}
    \caption{Reasoning results of the real-world motion video.}
    \label{fig:real_motion}
\end{figure*}

To evaluate the physical insights obtained by VER from real-world motion, we extract a video of vertical projectile motion from a public video website\footnote{https://www.youtube.com/watch?v=5GiCJ5Bjupw} as additional data. Compared to synthetic videos, real data has a more complex and noisy background, which poses challenges for identifying physical coordinates. We extract a complete sequence of the projectile motion from the original video and filter out blank frames to obtain a continuous set of 27 frames. As shown in Figure~\ref{fig:real_motion}, the experimental results demonstrate that VER accurately inferred the formula for vertical projectile motion.
Notably, SAM is unable to track the motion coordinates of the moving object, and thus can not derive its motion equations.

\begin{table*}[!t]
\renewcommand{\arraystretch}{1.}
\centering
\resizebox{1.0\linewidth}{!}{%
\begin{tabular}{cccccccccccc}
\toprule
 & \multicolumn{6}{c}{Pixel coordinate systems} & \multicolumn{5}{c}{Latent coordinate systems} \\
 & Linear & Cubic & Circular & VDP & Glider & Exp & LO & Bruss & Water & Video1 & Video2 \\
\midrule
Resolution & $500 \times 500$ & $500 \times 500$ & $500 \times 500$ & $500 \times 500$ & $500 \times 500$ & $500 \times 500$ & $100 \times 100$ & $64 \times 64$ & $128 \times 128$ & $128 \times 128$ & $128 \times 128$ \\
Frame & 200 & 200 & 200 & 200 & 200 & 200 & 1,000 & 1,000 & 1,000 & 310 & 420 \\
\bottomrule
\end{tabular}%
}
\caption{Descriptions of dataset size.}
\label{tab:dataset}
\end{table*}

\subsection{Adaptive regularization coefficient} \label{app:adaptive_eta}

The regularization coefficient $\eta$ in Equation~\ref{loss:sindy} constrains the length of the discovered equations.
Larger values enhance sparsity constraints, while smaller values explore complex equations. We add $\eta$ to MLLM's sample tuple, enabling it to select both the dictionary and based on past experience. Allowing MLLM to simultaneously adjust the dictionary and regularization coefficient is indeed a more flexible approach. Experimental results (Table~\ref{tab:eta_table}) show that adaptive selection yields significantly better long-term predictions for systems with higher-order terms.

\begin{table}[!ht]
\renewcommand{\arraystretch}{1.}
\centering
\resizebox{\linewidth}{!}{%
\begin{tabular}{ccccccc}
\toprule
 \multirow{2}{*}{Case} & \multirow{2}{*}{Method} & \multicolumn{3}{c}{Equation} & \multicolumn{2}{c}{Prediction} \\
 & & \makecell{Terms \\ Found} & \makecell{False \\ Positives} & $R^2$ & $R^2@100$ & $R^2@1000$ \\
\midrule
 \multirow{2}{*}{Linear} & VER & Yes & $0.38\pm0.22$ & $\mathbf{9.59\pm0.25}$ & $9.68\pm0.20$ & $9.62\pm0.23$ \\
 & $\eta$-VER & Yes & $\mathbf{0.35\pm0.19}$ & $9.54\pm0.20$ & $\mathbf{9.70\pm0.18}$ & $\mathbf{9.64\pm0.22}$ \\
\midrule
 \multirow{2}{*}{Cubic} & VER & Yes & $1.41\pm0.73$ & $9.53\pm0.32$ & $7.96\pm1.60$ & $6.16\pm2.78$ \\
 & $\eta$-VER & Yes & $\mathbf{1.07\pm0.52}$ & $\mathbf{9.61\pm0.26}$ & $\mathbf{8.45\pm0.98}$ & $\mathbf{7.23\pm2.05}$ \\
\midrule
 \multirow{2}{*}{Circular} & VER & Yes & \textbf{0} & $9.91\pm0.07$ & $\mathbf{9.99\pm0.00}$ & $9.67\pm0.45$ \\
 & $\eta$-VER & Yes & \textbf{0} & $\mathbf{9.92\pm0.06}$ & $9.97\pm0.04$ & $\mathbf{9.70\pm0.42}$ \\
\midrule
 \multirow{2}{*}{VDP} & VER & Yes & $0.80\pm0.40$ & $\mathbf{9.69\pm0.08}$ & $9.35\pm0.19$ & $7.28\pm0.72$ \\
 & $\eta$-VER & Yes & $\mathbf{0.51\pm0.27}$ & $9.45\pm0.05$ & $\mathbf{9.52\pm0.11}$ & $\mathbf{8.31\pm0.60}$ \\
\bottomrule
\end{tabular}%
}
\caption{Average performance of the adaptive regularization coefficient $\eta$-VER in the pixel coordinate systems. The magnitude of $R^2$ is $10^{-1}$.}
\label{tab:eta_table}
\end{table}

\subsection{Bayesian hyperparameter search} \label{app:bayes_ablation}

In the VER framework, the Savitzky-Golay filter and the symbolic dictionary of SINDy involve parameter selection. While we design visual enhancement prompts and the empirical pool for MLLM to unlock and leverage its vast pre-trained knowledge, an alternative option is Bayesian optimization. 
Here, we provide a comparison with Bayesian hyperparameter search to verify the necessity of the MLLM component, including: 1) \textit{Bayes-SINDy}: Replaces MLLM-based SINDy dictionary selection with Bayesian search (using scikit-optimize). The evaluation metric follows Equation~\ref{loss:sindy}; 2) \textit{Bayes-smooth}: Uses Bayesian search to select filter parameters, evaluated by the L1 error to the original trajectory; 3) \textit{Bayes-dim}: Replaces MLLM-based dimension selection with Bayesian search, evaluated by L2 reconstruction error on the validation set.

The results (Table~\ref{tab:ablation_bayes}) show that Bayesian search performs worse than VER. One possible explanation is that Bayesian optimization relies solely on quantitative scores, while VER also considers qualitative aspects such as equation length and visual smoothness. The lack of well-defined quantitative metrics makes this problem particularly challenging. For example, a larger symbolic dictionary may reduce fitting error but lead to overfitting. Equation complexity and fitting accuracy are often in tension, and our MLLM component is designed to balance them automatically.

\begin{table}[!t]
\renewcommand{\arraystretch}{1.0}
\centering
\resizebox{\linewidth}{!}{%
\begin{tabular}{ccccccc}
\toprule
 \multirow{2}{*}{Case} & \multirow{2}{*}{Method} & \multicolumn{3}{c}{Equation} & \multicolumn{2}{c}{Prediction} \\
 & & \makecell{Terms \\ Found} & \makecell{False \\ Positives} & $R^2$ & $R^2@100$ & $R^2@1000$ \\
\midrule
 \multirow{3}{*}{Linear} & Bayes-SINDy & Yes & $0.49\pm0.18$ & $9.37\pm0.41$ & $9.42\pm0.35$ & $9.08\pm0.47$ \\
 & Bayes-smooth & Yes & $2.51\pm1.02$ & $8.71\pm0.44$ & $8.56\pm0.54$ & $4.80\pm1.22$ \\
 & VER & Yes & $\mathbf{0.38\pm0.22}$ & $\mathbf{9.59\pm0.25}$ & $\mathbf{9.68\pm0.20}$ & $\mathbf{9.62\pm0.23}$ \\
\midrule
 \multirow{3}{*}{Cubic} & Bayes-SINDy & Yes & $3.64\pm1.39$ & $9.09\pm0.20$ & $7.51\pm1.48$ & $3.63\pm2.11$ \\
 & Bayes-smooth & Yes & $4.36\pm1.96$ & $7.72\pm1.04$ & $7.02\pm1.69$ & $2.94\pm1.86$ \\
 & VER & Yes & $\mathbf{1.41\pm0.73}$ & $\mathbf{9.53\pm0.32}$ & $\mathbf{7.96\pm1.60}$ & $\mathbf{6.16\pm2.78}$ \\
\midrule
 \multirow{3}{*}{Glider} & Bayes-SINDy & Yes & $1.05\pm0.35$ & $\mathbf{9.90\pm0.23}$ & $9.51\pm0.14$ & $8.33\pm0.66$ \\
 & Bayes-smooth & Yes & $1.15\pm0.30$ & $9.15\pm0.14$ & $9.29\pm0.09$ & $7.89\pm0.26$ \\
 & VER & Yes & $\mathbf{0.84\pm0.17}$ & $9.78\pm0.08$ & $\mathbf{9.90\pm0.07}$ & $\mathbf{9.23\pm0.18}$ \\
\bottomrule
\end{tabular}%
}
\caption{Ablation studies on linear and cubic systems. Bayesian hyperparameter search replaces MLLM in selecting the SINDy dictionary (Bayes-SINDy) and smoothing parameters (Bayes-smooth). The magnitude of $R^2$ is $10^{-1}$.}
\label{tab:ablation_bayes}
\end{table}

\subsection{Time cost}

\begin{figure}
    \centering
    \includegraphics[width=0.5\linewidth]{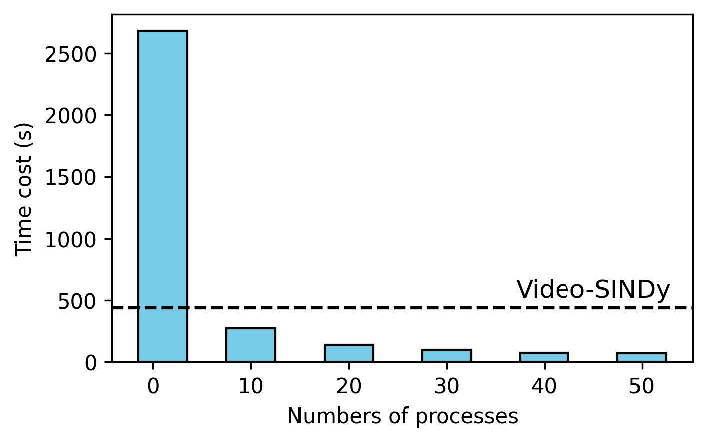}
    \caption{Time cost comparasion.}
    \label{fig:time}
\end{figure}

Our approach is zero-shot, in contrast to our baseline methods that rely on case-specific tuning. Although GPT-4o introduces some network latency, which impacts our test-time comparison, our method supports parallel execution, enabling significant speed-ups through multi-process parallelism.
We test time costs with varying numbers of processes $P$ and provide a comparison with the baseline Video-SINDy, which is trained on a single RTX 4060ti GPU with Batchsize=200. The results (Figure~\ref{fig:time}) show that VER, benefiting from its parallelization advantage, is significantly faster.

\section{Supplementary Materials}

Here we provide detailed algorithms, prompts, and output examples for the methodology section.

\begin{algorithm}[!ht]\label{algo:detect_coord}
   \caption{Detection CoT for Pixel Coordinate}
   \label{alg:example}
\begin{algorithmic}
   \STATE {\bfseries Input:} $q=\{x_l,x_i || x_l \in \mathbb{L}, x_i \in \mathbb{I}\}$; \ $\mathrm{DetTool}$; \ $\mathrm{MLLM}$; \ $\mathrm{Filter}$
   \STATE {\small {\textcolor{gray}{\# Initialize an empty coordinate list and filter hyperparameters}}}
   \STATE $Z \leftarrow$ InitCoord()
   \STATE $isSmooth, h, p \leftarrow true, h_0, p_0$
   \STATE {\small {\textcolor{gray}{\# Serially infer the physical coordinate sequence}}}
   \FOR{$i=1$ {\bfseries to} $n$}
   \STATE $q_i \leftarrow \mathrm{DetTool}(q_i)$
   \STATE $y_i \leftarrow \mathrm{MLLM}(q_i)$
   \STATE $Z \leftarrow Z \cup \{e(y_i)\}$
   \ENDFOR
   \STATE {\small {\textcolor{gray}{\# Adaptive smoothing filter}}}
   \WHILE{$isSmooth$}
   \STATE {\small {\textcolor{gray}{\# $h$ is visual and textual prompt for smoothing results}}}
   \STATE $Z, h \leftarrow \mathrm{Filter}(Z,h,p)$
   \STATE $isSmooth \leftarrow \mathrm{MLLM}(h)$
   \ENDWHILE
   \STATE {\bfseries Output:} $Z$
   
\end{algorithmic}
\end{algorithm}

\begin{table}[!ht]
\renewcommand{\arraystretch}{1.0}
\centering
\resizebox{\linewidth}{!}{%
\begin{tabular}{cccccc}
\toprule
 \multirow{2}{*}{Case} & \multirow{2}{*}{Method} & \multicolumn{4}{c}{Noise strength} \\
 & & $R^2@0.0$ & $R^2@0.1$ & $R^2@0.2$ & $R^2@0.3$ \\
\midrule
 \multirow{2}{*}{LO} & AE-SINDy & $9.41\pm0.04$ & $9.06\pm0.11$ & $8.37\pm0.18$ & $7.94\pm0.47$ \\
 & VER & $\mathbf{9.69\pm0.06}$ & $\mathbf{9.57\pm0.05}$ & $\mathbf{9.08\pm0.20}$ & $\mathbf{8.26\pm0.32}$ \\
\midrule
 \multirow{2}{*}{Bruss} & AE-SINDy & $\mathbf{8.62\pm0.23}$ & $7.72\pm0.44$ & $6.91\pm0.81$ & $6.63\pm0.92$ \\
 & VER & $8.50\pm0.07$ & $\mathbf{8.32\pm0.18}$ & $\mathbf{7.68\pm0.61}$ & $\mathbf{6.96\pm0.82}$ \\
\midrule
 \multirow{2}{*}{Water} & AE-SINDy & $9.64\pm0.08$ & $9.00\pm0.35$ & $8.07\pm0.55$ & $7.03\pm1.30$ \\
 & VER & $\mathbf{9.72\pm0.10}$ & $\mathbf{9.66\pm0.28}$ & $\mathbf{8.61\pm0.60}$ & $\mathbf{7.67\pm0.95}$ \\
\bottomrule
\end{tabular}%
}
\caption{The average performance of the latent coordinate system under different noise strengths, run more than 10 times with different random seeds. $R^2@\sigma$ is the $R^2$ metric, with a magnitude of $10^{-1}$, of the identified equation when the noise strength is $\sigma$.}
\label{tab:noise}
\end{table}

\end{document}